\documentclass[aps,prl,twocolumn,showpacs,groupedaddress]{revtex4}
\usepackage{natbib}
\usepackage{amsmath}
\usepackage{amssymb}
\usepackage{amsthm}
\usepackage{graphicx}
\usepackage{graphics}
\usepackage{bm}
\usepackage{hyperref}
\usepackage{epsfig}
\usepackage{dcolumn}

\def\qpc{\textsf{QPC}~}
\def\qpcd{\textsf{QPC}}

\def\cftd{\textsf{CFT}}

\def\wbs{\textsf{WB}~}
\def\wbsd{\textsf{WB}}
\def\sbs{\textsf{SB}~}
\def\sbsd{\textsf{SB}}

\begin{document}

\title{Probing the neutral edge modes in transport across
a point contact via thermal effects in the Read-Rezayi non-abelian
quantum Hall states}

\author{Eytan Grosfeld$^{1}$ and Sourin Das$^{2}$}
\affiliation{$^{1}$ Department of Physics, University of Illinois,
1110 W. Green St., Urbana IL 61801-3080, U.S.A. \\
$^{2}$ Center for High Energy Physics, Indian Institute of Science,
Bangalore 560 012, India}

\date{\today}

\begin{abstract}

Non-abelian quantum Hall states are
characterized by the simultaneous appearance of charge and neutral
gapless edge modes, with the structure of the latter being intricately related to the
existence of bulk quasi-particle excitations obeying non-abelian
statistics. In general, it is hard to probe the neutral modes
in charge transport measurements and a thermal transport measurement
seems to be inevitable. Here we propose a setup which can get around
this problem by having two point contacts in series separated by
a distance set by the thermal equilibration length of the charge mode. We show
that by using the first point contact as a heating device, the excess charge noise measured at the second point
contact carries a non-trivial signature of the presence of the neutral
mode hence leading to its indirect detection. We also obtain explicit
expressions for the thermal conductance and corresponding Lorentz number
for transport across a quantum point contact between two edges held at
different temperatures and chemical potentials.
\end{abstract}


\maketitle The Read-Rezayi non-abelian quantum Hall states are
currently the leading candidates for ground state wave-functions
describing certain Hall plateaus appearing on the first Landau
level. The first state of this series coincides with the Pfaffian
state suggested by Moore and Read in 1991 \cite{MooreRead} which
accounts for the plateau at filling factor $\nu=2+1/2$
\cite{ReadGreen}. The rest of the series describes other observed
plateaus at filling factors $\nu=2+2/(k+2)$ ($k=3,4,\ldots)$
\cite{ReadRezayi}. 

The edge states of the Read-Rezayi quantum Hall
states are described by two independent $1+1$ dimensional field
theories~\cite{milovanovic}: a bosonic field theory carrying
$\mathbb{U}(1)$ charge and a charge-neutral field theory belonging
to a class of conformal field theories (\cftd) known as
parafermions~\cite{ZamolodchikovFateev}. These two theories together
are responsible for the low energy transport properties of these
quantum Hall states. The structure of the neutral edge is set by the
bulk theory, and it directly reflects the non-abelian properties of
the quasi-particles \cite{FradkinNayakTsvelikWilczek}. Hence, it is
the properties of this edge that one aims to probe in experimental
work when trying to detect signatures of the non-abelian states.

Recently there has been a considerable interest in understanding
charge transport across a single quantum point contact
(\qpcd)~\cite{FendleyFisherNayak} or multiple \qpcd's
(interferometer geometries)~\cite{SternHalperin,
BondersonKitaevKirill,BondersonKirillSlingerland,ChungStone,Ilan}.
These studies mainly focus on tunneling of quasi-particles or
electrons across a \qpc in the weak-backscattering
 (\wbsd) limit or in the strong-backscattering (\sbsd) limit, when a
voltage is applied across the \qpcd. Charged quasi-particles or
electrons carry the neutral part of the excitation along with them as
they tunnel across a \qpcd. The electric current therefore possesses
some of the properties of the neutral field theory. However,
past theoretical developments indicate that charge transport
measurements in a single \qpc cannot confirm the existence of
neutral edge modes, and the minimal requirement for probing the
non-abelian nature of the state is an interferometer geometry which
is yet to be realized experimentally. On the other hand, there has
not been much progress in exploring possibilities for the detection
of neutral-edge modes keeping limitations of present day experiments
in mind. 

In this letter, we propose a scenario to probe the neutral
edge mode directly through thermal current measurements, and provide
the theoretical framework for calculating observables related to
such an experiment. In sharp contrast to standard  charge transport
measurements, a temperature gradient directly couples to the neutral
mode, and observables such as the thermal conductance will therefore
reflect the presence of the neutral mode. But in view of the fact
that controlled thermal conductance measurements are beyond the
scope of present day experiments, we propose a setup which does not
require any external heating devices, but at the same time
indirectly probes the presence of neutral mode via thermal effects.
We start with a brief introduction to the physics of the thermal
current carried by the edge uninterrupted by a \qpcd, and find the
$2$-terminal thermal Hall conductance and Lorentz number
(Eq.~(\ref{eq:pure-lorentz-number})). We then perturbatively
calculate the charge and energy currents associated with tunneling
of particles of scaling dimension $h$ across a \qpc to leading order
in tunneling amplitude and provide an expression for it as function
of the voltage bias, $V$, and temperature bias, $\Delta T$ (see
Eq.~(\ref{eq:thermal-conductance})). The associated Lorentz number
is calculated as well (Eq.~(\ref{eq:lorentz-number})). We then
proceed to elaborate on the experimental possibility of using a \qpc as
a heating device  for the charge edge incident on the \qpc leaving
behind the co-propagating neutral edge at its initial temperature.
This is naturally achieved in the presence of a voltage bias as the applied voltage almost entirely drops between the
counter-propagating charge edge modes at the \qpcd. 
The out-of-equilibrium charge mode
coming out of the \qpc region undergoes self-equilibration via the
intra-edge Coulomb interaction, hence attaining a new temperature
which is different from the temperature of the edge state incident
on the \qpc due to heating of the charge mode at a rate given by
$P=I V$ ($I$ is the tunneling current and $V$ is the voltage drop
across the \qpcd). This results in a sustained temperature
difference between the co-propagating charge and neutral edge modes
coming out of \qpcd. We show that this temperature difference can be
detected via a noise measurement by invoking a second \qpc in
series with the first one (Eqs.~(\ref{eq:noise}),
(\ref{eq:effective-charge})). As the origin of the temperature
difference is solely this fact that one of the edges is
charge-neutral, its detection can confirm the presence of the
neutral mode without resorting to a thermal conductance measurement.

\paragraph{Thermal Hall conductance for uninterrupted edge.-} The
thermal current carried along the edge is given by
\cite{Cappelli,KaneFisher-luttinger,KaneFisher-hierarchy}
\begin{eqnarray}
    I_Q = c \dfrac{\pi^2}{6} \dfrac{k_B^2}{2\pi} T^2,
\end{eqnarray}
where $T$ is the temperature of the edge and $c=c_n+c_c$ is
the total central charge, with $c_n$ ($c_c$) being the central
charge for the neutral (charge-carrying) edge theory. We have set
$\hbar=1$ throughout the letter. For the Laughlin states there are
no neutral modes, hence $c=c_c=1$. For the Read-Rezayi series
$c_n=(2k-2)/(k+2)$, $c_c=2+1$ ($k=2,3,4,\ldots$) where the $2$ in
$c_c$ accounts for the two integer Hall edges of the $\nu=2$ state. The
corresponding two-terminal thermal Hall conductance is
\begin{eqnarray}
    \label{eq:pure-thermal-conductance}
    K_H = c \dfrac{\pi^2}{3} \dfrac{k_B^2}{2\pi} T.
\end{eqnarray}
Plugging the value of the electric Hall conductance $G_H=\nu
e^2/2\pi$ into the definition of the Lorentz number we get
\begin{eqnarray}
    \label{eq:pure-lorentz-number}
    L=\dfrac{K_H}{G_H T} &=& \dfrac{c}{\nu}
    \dfrac{\pi^2}{3}\dfrac{k_B^2}{e^2}=\dfrac{c}{\nu}L_0,
\end{eqnarray}
where $L_0=\frac{\pi^2}{3}\frac{k_B^2}{e^2}$ is the Lorentz number
for a free electron gas. For the Laughlin series it is given by
$L_{\mathrm{LA}}=L_0/\nu$, where $\nu=1/(2p+1)$ ($p=0,1,2,\ldots$)
\cite{KaneFisher-luttinger}. For the Read-Rezayi series,
$\nu=2+k/(k+2)$ and $c=2+3k/(k+2)$, we get
\begin{eqnarray}
    L_{\mathrm{RR}}=\dfrac{5k+4}{3k+4} L_0.
\end{eqnarray}
It might be of interest later to consider the Lorentz number for
the partially filled upper Landau level, $\nu=k/(k+2)$, leaving behind
the contribution of the integer Hall edge states ($\nu=2$). The Lorentz number
is then $L_{\mathrm{RR}}=L^n_{\mathrm{RR}}+L^c_{\mathrm{RR}}=3L_0$
which is independent of $k$, with the neutral and charge
contributions being
    $L^n_{\mathrm{RR}}=\left(\frac{2k-2}{k}\right)L_{0}~,
    L^c_{\mathrm{RR}}=\left(\frac{k+2}{k}\right)L_0$
respectively. This Lorentz number can be measured in
a geometry \cite{drs5/2} where the $\nu=2$ edges are fully
back-scattered while transmitting the fractional edges perfectly at
a \qpcd.

\paragraph{Perturbative calculation of tunneling thermal currents
across a point contact.-} We write the Hamiltonian for the edges
corresponding to the partially filled upper Landau level as
\begin{eqnarray}
    H = \sum_{i=L,R}\left(H^i_{n}+H^i_{c}\right)+H_{\mathrm{tun}}.
\end{eqnarray}
Here $H_{c}^{L/R}$ describe the charged degrees of freedom in the
absence of tunneling,
\begin{eqnarray}
    H^i_{c}=\dfrac{v_c}{4\pi}\int dx
    \left(\partial\phi_i(x)\right)^2,
\end{eqnarray}
where $v_c$ is the charge edge velocity, and $H^{L/R}_n$ describe the
neutral degrees of freedom ($L/R$ correspond to left/right
movers). Although the latter
Hamiltonian cannot be written explicitly for all the Read-Rezayi
states, it acquires an exact meaning within the formalism of
conformal field theory as the zeroth mode in the Laurent expansion
of the energy-momentum tensor (see, \textit{e.g.} \cite{CFTbook}). Finally, the tunneling
term corresponding to \qpc which mixes only the edges corresponding
to the upper Landau level is given by
\begin{eqnarray}
    H_{\mathrm{tun}} = \lambda \Phi_L^\dag \Phi_R+\mathrm{h.c.},
\end{eqnarray}
where $\Phi$ is the most relevant operator for tunneling, as
dictated by the experimental scenario. We can decompose the operator
into its neutral and charge components, as $\Phi=\Phi_n \Phi_c$,
where $\Phi_{n/c}$ has scaling dimension $h_{n/c}$ respectively.
Hence the scaling dimension of $\Phi$ will be $h=h_n+h_c$ (Table \ref{table:scaling-dimensions}). We next
turn to calculate the electric current and thermal current to
leading order in $\lambda$ in perturbation theory, and then
calculate corresponding Lorentz numbers. We briefly discuss the
electric current, closely following Chamon {\it{et al.}}
\cite{chamon}, extending the formalism to the case that a
temperature difference is maintained between the edges. The electric
current is defined as the rate of particle (electron/quasi-particle)
transfer between the two edges times the effective charge
\begin{eqnarray}
    I &=& -e^{\star}\dfrac{i}{2} \left[N^L-N^R,H\right].
\end{eqnarray}
Here $N_{L/R}$ correspond to the number of
electrons on the left (right) moving edge and $e^\star$ is the
charge of tunneling particle (\textit{i.e.} $e^{\star}$ stands for electron charge in the \sbs limit and quasi-particle charge in the \wbs limit). To lowest order in the tunneling, the
electric current through a single \qpc is given by the following
expression \cite{chamon}
\begin{eqnarray}
    \langle I_h\rangle = e^{\star} |\lambda|^2
    \left[P_h(\omega_0,T,\Delta T)-P_h(-\omega_0,T,\Delta T)\right].
    \end{eqnarray}
Here $\omega_0=e^\star V$ is the Josephson frequency, and $P_h$ is
defined as
\begin{eqnarray}
    P_h(\omega,T,\Delta T) = \int_{-\infty}^{\infty} dt  e^{i\omega
    t}G_{T+\frac{\Delta T}{2}}(t)G_{T-\frac{\Delta T}{2}}(t),
\end{eqnarray}
with the finite temperature Green's function being $
    G_T(t)=\left[{\pi T}/{\sin\{\pi T\left[\delta+i t\right]\}}\right]^{2h}$.
Clearly, by expanding $P_h$ in $\Delta T$, there is no contribution
linear in $\Delta T$. Hence to order $\Delta T$ the electrical current
written in terms of the conformal dimension of the tunneling
particle is \cite{chamon}
\begin{eqnarray}
    \label{eq:current}
    \langle I_h \rangle &=&
   \dfrac{e^\star|\lambda|^2}{(2\pi T)^{1-4h}}  {\cal B} \left[2h-\frac{i\omega_0}{2\pi T},2h+
    \frac{i\omega_0}
    {2\pi T} \right]
     \sinh\left(\frac{\omega_0}{2 T}\right).\nonumber\\
\end{eqnarray}
Here ${\cal B}(x,y)=\Gamma(x)\Gamma(y)/\Gamma(x+y)$ is the beta
function. Due to linearity of spectrum of fundamental excitations in
the neutral and the charge sector, the electric current vanishes in the zero voltage bias
limit even if two edges are
maintained at two different temperatures. We now repeat this
calculation for the heat current. While only the charge-carrying
mode contributes to the electric current, here the neutral mode
contributes as well. In the following, we will write the tunneling
heat current as a sum of two terms, $I_Q=I_{Q,n}+I_{Q,c}$, where the
indices $n$ and $c$ refer to the neutral mode and charge mode
contributions respectively. We define $I_{Q,n/c}$ as the rate of
energy transfer between the left and right moving edges,
\begin{eqnarray}
    I_{Q,n/c}=&-\dfrac{i}{2} [H^L_{n/c}-H^R_{n/c},H ].
\end{eqnarray}
The expectation values of $I_{Q,n/c}$ can be written as
\begin{eqnarray}
 \label{eq:fraction}
\langle I_{Q,n}\rangle = \frac{h_n}{h_n+h_c}\tilde{I}_{Q}~; \,
    \langle I_{Q,c}\rangle =\frac{h_c}{h_n+h_c}\tilde{I}_{Q}+
    V \langle I_h \rangle.
\end{eqnarray}
Note that $\tilde{I}_Q$ goes to zero as $\Delta T\to 0$, and is
given by
\begin{eqnarray}
  &&  \tilde{I}_Q
    = -i\frac{|\lambda|^2}{2}\big[
    Q(\omega_0,T,\Delta T)-Q(\omega_0,T,-\Delta T)
\nonumber\\
  && \quad \quad + Q(-\omega_0,T,\Delta T)-Q(-\omega_0,T,-\Delta
    T)\big]~;  ~~ {\mathrm{with}}
      \nonumber\\
%
   && \!\!\!\! Q(\omega,T,\Delta T) = \int_{-\infty}^{\infty}dt\, e^{i\omega
    t} G_{T+\frac{\Delta T}{2}}(t)\partial_t G_{T-\frac{\Delta T}{2}}(t).
\end{eqnarray}
Therefore, in response to a temperature difference, 
for electron/quasi-particle operators which are composed of
fields from both charged and neutral modes, the amount of energy
carried by them while tunneling splits among the two modes
proportionally to the percentage scaling dimension
($\frac{h_{c/n}}{h_c+h_n}$) of each constituent field.

To first order in the temperature difference $\Delta T$, we can
write $\tilde{I}_Q=K\Delta T$, where $K$ is the longitudinal thermal
conductance. Using the expressions above, the thermal conductance
expressed in terms of the electric current, Eq.~(\ref{eq:current})
is given by
\begin{eqnarray}
    \label{eq:thermal-conductance}
    K_h=\frac{1}{2 T e^\star}\left(2 h \partial_{\omega_0} \langle I_{h+1/2}
     \rangle-\frac{1}{2}\omega_0 \langle I_h \rangle\right).
\end{eqnarray}
The corresponding Lorenz number is
\begin{eqnarray}
    \label{eq:lorentz-number}
    L_h=\frac{K_h}{T G_h}=\frac{12 h^2}{1+4
    h}\frac{e^2}{(e^\star)^2}L_0.
\end{eqnarray}
Here $G_h$ is the linear conductance defined as $\langle
I_h\rangle /V$ in the $V\rightarrow 0$ limit. Plugging the
scaling dimension for an electron in the Laughlin states,
$h=1/2\nu$, the Lorentz number coincides with the result given by
Kane and Fisher \cite{KaneFisher-luttinger}. This is one of the
central results of this letter. From Eq.~(\ref{eq:fraction}) the
charged and neutral contributions to the Lorentz number are
$L_n=\frac{h_n}{h}L_h,L_c=\frac{h_c}{h}L_h$.
Therefore for tunneling thermal current corresponding to electron
tunneling between edge states (which is the case for almost closed
\qpcd) of the upper Landau level, the energy splits among the two
modes as ${\frac {h_n}{h_c}}=2{\frac{k-1}{k+2}}$ which is identical
to that of the uninterrupted edge, $\frac{c_n}{c_c}
=2\frac{k-1}{k+2}$. However this may be special to the Read-Rezayi
series.
\begin{table}
\begin{tabular}{|l|c|c|c|c|}
\hline
 & $e^\star/e$ & $h_c$ & $h_n$ & $h$\\
\hline
 & & & & \\
Electron & $1$ & $\frac{k+2}{2k}$ & $\frac{k-1}{k}$ & $\frac{3}{2}$\\
 & & & & \\
\hline
 & & & & \\
Quasi-particle & $\frac{1}{k+2}$ & $\frac{1}{2k(k+2)}$ &
$\frac{k-1}{2k(k+2)}$
& $\frac{1}{2(k+2)}$\\
 & & & & \\
\hline
\end{tabular}
\caption{The effective charge and scaling dimensions of the electron
and quasi-particle in the $\nu=2+k/(k+2)$ states. Here $h_n$, $h_c$
and $h$ are the neutral, charge and total scaling dimensions
respectively.}
\label{table:scaling-dimensions}
\end{table}

\begin{figure}
\includegraphics[scale=0.25]{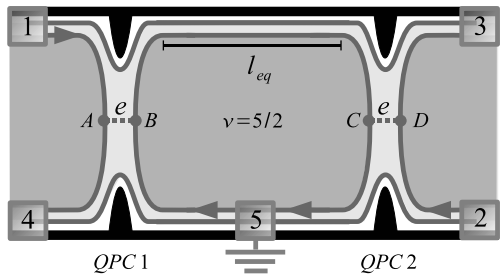}
\caption{\qpcd-$1$ is used as a heating device for the charged mode.
After a distance $\ell_{\mathrm{eq}}$, a temperature imbalance is
established between the neutral and charge edge (at point $B$:
$T_n=T_c=T$. At point $C$: $T_n=T$, $T_c=T+\Delta T_{nc}$). The fully
transmitting edge at both \qpcd s is the $\nu=2$ edge.}
\label{fig:setup}
\end{figure}

\paragraph{Proposed Geometry.-}
We now consider the experimental proposal depicted in
Fig.~\ref{fig:setup}. Consider a situation where a voltage bias is
applied at contact-$1$ with respect to the grounded contact-$5$
driving a current at \qpcd-$1$. Here we assume that the edge state
emanating out of contacts $1,2,5$ are at equilibrium and at a common
temperature $T$. The current and the associated energy injected into
the edge starting at point $B$ (see Fig.~\ref{fig:setup}) and
heading towards \qpcd-$2$ due to tunneling of electrons at \qpcd-$1$
is solely pumped into the charge mode (up to finite size
contributions to the neutral mode which go to zero for large
system sizes). Now, if the charge mode equilibrates via intra-edge
Coulomb interaction, then, under certain conditions on which we
elaborate below, it can maintain a temperature different than the
co-propagating neutral mode temperature. There are three relevant
time scales which we need to worry about: the time required for the
charge bosonic edge to reach equilibrium $\tau_{c}$, the time
required for the neutral edge to reach equilibrium $\tau_n$, and the
time required for the two edges to reach mutual equilibrium
$\tau_{nc}$. In the case that $\tau_{nc}\gg \tau_c$, as energy is
pumped only into the charged mode, the two edges can indeed maintain
a different temperatures with $T_c>T$ and $T_n=T$ after the charge
edge has undergone self-equilibration. The microscopic mechanisms
required to estimate these three timescales are not understood at
this stage. However, on general grounds, we expect that $\tau_{nc}$
will be much larger than $\tau_c$, as the charge bosonic edge can
interact via the Coulomb interaction, while the neutral edge is more
inert.
%
%
The pumping of charge and energy into the edge emanating out of
contact-$5$ due to tunneling events at \qpcd-$1$ will result in
increase of both temperature and voltage of this edge after it
undergoes equilibration. Assuming both charge and energy currents are
conserved at \qpcd-$1$, we find
\begin{eqnarray} \label{eq:deltaVT}
    \Delta V = \langle I\rangle/G_H, \;\;\;\;\; \Delta T_{nc}=
\langle I \rangle V/K_{H}.
\end{eqnarray}
One of the most useful 'thermometers' available to experimentalists
to measure the electronic temperature on the edge is through shot
noise measurements. Here we argue that shot noise measurements can
be used to detect the temperature \textit{difference} $\Delta
T_{nc}$ as well. The purpose of the second \qpc in
Fig.~\ref{fig:setup} placed at a distance larger than the charge
bosonic edge equilibration length is to generate noise. Following
Ref.~\onlinecite{nayak} the noise in the present context either in
the weak or the strong backscattering limit can be evaluated to be
\begin{eqnarray}
    \label{eq:noise}
    S=2 \tilde{e}^{\star}(\omega_0,T,\Delta T_{nc})\langle e^\star I_h(\omega_0,T)
    \rangle \coth\left(\frac{\omega_0}{2 T}\right),
\end{eqnarray}
with the new effective charge being a function of bias, the
temperature and the temperature difference $\Delta T_{nc}$:
\begin{eqnarray}
    \label{eq:effective-charge}
    \tilde{e}^{\star}(\omega_0,T,\Delta T_{nc})=e^\star \left(1+\frac{\Delta T_{nc}}
    {T}\frac{h_c}{2 h}\frac{\omega_0/T}{\sinh(\omega_0/T)}\right).
\end{eqnarray}
This $\Delta T_{nc}$ dependent ${e}^{\star}$ is an indirect probe
for detection of the neutral mode and in addition the coefficient
$h_c/2h$ which is smaller than $1/2$ also indicates the existence of
a neutral mode. It should be noted that $\omega_0$ represents the
voltage bias at \qpcd-$2$ which is controlled by the difference
between $\Delta V$ and the voltage applied at contact-$2$. Hence
carrying out a voltage sweep at contact-$2$ and
measuring the corresponding $e^\star$ via noise measurement at
terminal-$3$ can provide a direct check of our predictions. For our
proposal to work, the most crucial part is to obtain a reasonable
$\Delta T_{nc}$ for typical experimental accessible currents and
2DEG temperatures. Using Eq.~(\ref{eq:deltaVT}), for $T =
40\times 10^{-3}$~Kelvin, impinging current on \qpcd-$1$ to be $.2\times
10^9$ Amps., transmission of \qpc to be $1\%$  and assuming
terminal-$5$ be grounded we get $\Delta T_{nc}=3\times 10^{-3}$~Kelvin
which is well within resolution of present day experiments.

To conclude, we have perturbatively calculated the thermal
conductance and Lorentz number for tunneling of any excitation
(quasi-particles/electrons) between the edges of Read-Rezayi quantum
Hall states in terms of the scaling dimension of the tunneling
operator. We discuss the application of a \qpc as a heating
device and propose a setup where it can be used to detect the
presence of the neutral mode. We also provide an expression for tunneling
noise and the corresponding effective charge measured for edge
states where the charge and the neutral edges are at two different
temperatures.

We acknowledge support from Weizmann Institute where this
work was started. SD thanks J. Eisenstein for sharing his views on
using a \qpc as a heating device. EG thanks A. Stern and E. Fradkin for useful discussions. This work was supported in part by the ICMT (EG) and the DST project (SR/S2/CMP-27/2006) (SD).

\bibliographystyle{apsrev}
\bibliography{eytan-sourin-thermal-sub}

\end{document}